\documentclass[12pt,letterpaper]{article}

\usepackage{scicite}
\usepackage[bottom]{footmisc}

\usepackage{fullpage}
\usepackage{amsmath}
\usepackage{amssymb}
\usepackage{amsthm}
\usepackage{textcomp}
\usepackage{fancybox}
\usepackage{subfigure}
\usepackage{sgame}

\usepackage{url}
\usepackage{tikz}
\usetikzlibrary{arrows,decorations,decorations.shapes,backgrounds,shapes}

\newtheorem{theorem}{Theorem}
\newtheorem{lemma}{Lemma}
\newtheorem{proposition}{Proposition}
\newtheorem{corollary}{Corollary}

\newtheorem{definition}{Definition}

\newenvironment{sciabstract}{%
\begin{quote} \bf}
{\end{quote}}

\newcounter{lastnote}

\title{Time Critical Social Mobilization: The DARPA Network Challenge Winning Strategy\footnote{This work was performed under U.S. Air Force contract FA8721-05-C-0002. Opinions, interpretations, conclusions, and recommendations are not necessarily endorsed by the U.S. Government.}}

\author
{Galen Pickard,$^{1,2}$ Iyad Rahwan,$^{3,1}$ Wei Pan,$^1$ Manuel Cebrian,$^1$\\ Riley Crane,$^1$ Anmol Madan,$^1$ Alex Pentland$^{1\dag}$\\
\\
\normalsize{$^{1}$Massachusetts Institute of Technology, USA}\\
\normalsize{$^{2}$Massachusetts Institute of Technology Lincoln Laboratory, USA}\\
\normalsize{$^{3}$Masdar Institute of Science \& Technology, UAE}\\
\\
\normalsize{$^\dag$Corresponding author; E-mail:  \url{pentland@mit.edu}}
}

\begin{document}

\baselineskip24pt

\maketitle

\begin{sciabstract}
It is now commonplace to see the Web as a platform that can harness the collective abilities of large numbers of people to accomplish tasks with unprecedented speed, accuracy and scale \cite{hand:2010}. To push this idea to its limit, DARPA launched its Network Challenge, which aimed to ``\textit{explore the roles the Internet and social networking play in the timely communication, wide-area team-building, and urgent mobilization required to solve broad-scope, time-critical problems}'' \cite{darpa:net:challenge}. The challenge required teams to provide coordinates of ten red weather balloons placed at different locations in the continental United States. This large-scale mobilization required the ability to spread information about the tasks widely and quickly, and to incentivize individuals to act. We report on the winning team's strategy, which utilized a novel recursive incentive mechanism to find all balloons in under nine hours. We analyze the theoretical properties of the mechanism, and present data about its performance in the challenge.
\end{sciabstract}

\section{Time-Critical Social Mobilization}

With the advent of communication technologies, and the Web in particular, we can now harness the collective abilities of large numbers of people to accomplish tasks with unprecedented speed, accuracy and scale. In popular culture and the business literature, this process has come to be known as \emph{crowdsourcing} \cite{howe:2009}.

In crowdsourcing, an interested party provides incentives for large groups of people to contribute to the completion of a task (or set of tasks). The nature of the tasks and the incentives vary substantially, ranging from monetary rewards, to entertainment, to social recognition. For example, a new breed of Web-based \emph{games with a purpose} provide hundreds of thousands of users with entertainment in exchange for completing highly complex tasks such as labeling tens of millions of images \cite{vonahn:2006} or predicting protein structures \cite{cooper:etal:2010}. Crowdsourcing markets, like Amazon's \emph{Mechanical Turk} \cite{pontin:2007}, allow anyone to post so-called Human Intelligence Tasks (HITs) that can be completed by people in exchange for payment. In \emph{prediction markets}, highly accurate aggregate predictions are accomplished by providing people with payments that depend on the accuracy of their individual predictions \cite{arrow:etal:2008}. And in Web sites relying on user-contributed content, such as YouTube, contributing users are motivated by the attention that their content receives from the community \cite{huberman:etal:2009}.

A particularly challenging class of crowdsourcing problems require not only the recruitment of a very large number of participants, but also extremely fast execution. Tasks that require this kind of \emph{time-critical social mobilization} include search-and-rescue operations in the aftermath of natural disasters, hunting down wanted outlaws on the run, reacting to health threats that need instant attention, or rallying  supporters to vote in a political campaign.

In time-critical social mobilization problems, it is often not practical, or even impossible, to create sufficient mobilization through mass media, due to the extremely high cost of reaching everybody, or due to severe infrastructure damage. In such cases, one has to resort to distributed modes of communication for information diffusion. For example, in the aftermath of Hurricane Katrina, amateur radio volunteers helped relay $911$ traffic for emergency dispatch services in areas that experienced severe communication infrastructure damage \cite{krakow:2005}.

Another common characteristic of these social mobilization problems is the presence of some sort of search process. For example, search may be conducted by members of the mobilized community for survivors after a natural disaster. Another kind of search attempts to identify individuals within the social network itself, such as finding a medical specialist to assist with a challenging injury in a natural disaster area.

There is growing literature on search in social networks. It has long been established that social networks are very effective at finding target individuals through short paths \cite{milgram:1967}, and various explanations of this phenomenon have been given \cite{kleinberg:2000,watts:etal:2002,adamic:adar:2005,rosvall:etal:2005}. 

However, it is important to recognize that the success of search in social mobilization requires individuals to be motivated to actually conduct the search, participate in the information diffusion, and so on. In other words, a key challenge in social mobilization is the \emph{incentive challenge}. Indeed, in an empirical study of search in a global social network, Dodds et al conclude that ``\textit{although global social networks are, in principle, searchable, actual success depends sensitively on individual incentives}'' \cite{dodds:etal:2003}. It has also been observed that the the success of crowdsourcing mechanisms, in general, can vary depending on the details of the financial incentive scheme in place \cite{mason:watts:2009}.

In summary, achieving time-critical, large-scale mobilization towards a problem requires two key ingredients: (a) the ability to spread information about the tasks widely and quickly, under constraints on the ability to broadcast such information; and (b) the provision of incentives for individuals to act, both towards the task \emph{and} towards the diffusion of information about it.

Recognizing the difficulty of time-critical social mobilization, the Defense Advanced Research Projects Agency (DARPA) announced the DARPA Network Challenge. The announcement, which coincided with the $40$th anniversary of the first remote log-in on the ARPA Net (considered the `birthday' of the Internet), was made at the University of California in Los Angeles on October $29$, $2009$. 

Through this challenge, DARPA aimed to ``\textit{explore the roles the Internet and social networking play in the timely communication, wide-area team-building, and urgent mobilization required to solve broad-scope, time-critical problems}'' \cite{darpa:net:challenge}. The challenge is to provide coordinates of ten red weather balloons placed at different locations in the continental United States. According to  DARPA, ``\textit{a senior analyst at the National Geospatial Intelligence Agency characterized the problem as impossible}'' by conventional intelligence gathering methods \cite{darpa:net:report}.

\section{The Recursive Incentive Mechanism}

According to the DARPA report, between $50$ and $100$ serious teams participated in the DARPA Network Challenge, from a total of $4,000$ teams \cite{darpa:net:report}. Moreover, approximately $350,000$ people participated in the DARPA Network Challenge in various ways, ranging from searching for balloons, to simply being aware of the challenge and willing to report a balloon if spotted. 

The MIT Team, which won the challenge \cite{darpa:winner}, completed the challenge in $8$ hours and $52$ minutes. In approximately $36$ hours prior to the beginning of the challenge, the MIT Team was able to recruit almost $4,400$ individuals through a recursive incentive mechanism.

%

The MIT Team's approach was based on the idea that achieving large-scale mobilization towards a task requires (a) diffusion of information about the tasks through social networks; and (b) provision of incentives for individuals to act, \emph{both} towards the task and towards the recruitment of other individuals.

We consider the MIT Team's approach to the DARPA Network Challenge to be an instance of a more general class of mechanisms for distributed task execution. We now define this class of mechanisms. But we first need to define the setting in which such mechanisms operate. We define a \emph{diffusion-based task environment} which consists of the following:
	$N = \{\alpha_1, \dots, \alpha_n \}$ is a set of \emph{agents};
	$E \subseteq N \times N$ is a set of \emph{edges} characterizing social relationships between agents;
	$\Psi = \{\psi_1, \dots, \psi_m \}$ is a set of \emph{tasks};
	$P : N \times \Psi \rightarrow [0,1]$ returns the \emph{success probability} of a given agent in executing a given task;
	$B \in \mathbb{R}$ be the \emph{budget} that can be spent by the mechanism.

In a diffusion-based task environment, unlike in traditional task allocation mechanisms (e.g. based on auctions), agents are \emph{not} aware of the tasks a priori. Instead, they \emph{become} aware of tasks as a result of either (1) being directly informed by the mechanism through advertising; or (2) being informed through recruitment by an acquaintance agent \cite{watts:peretti:2007}.


Another characteristic of diffusion-based task environments is that, when a task is completed, the mechanism is able to identify not only the agent who executed it, but also the information pathway that led to that agent learning about the task. The pathway leading to the successful completion of task $\psi_i$ is captured by the sequence $\mathcal{S}(\psi_i) = \langle a_1, \dots, a_r \rangle$ of unique agents, where $a_r$ is the agent who completed the task, $a_r$ was informed of the task by $a_{r-1}$ and so on up to agent $a_1$ who was initially informed of the task by the mechanism. By slightly overloading notation, let $|\mathcal{S}(\psi_i)|$ denote the length of the sequence (i.e. the number of agents in the chain), and let $\alpha_j \in \mathcal{S}(\psi_i)$ denote that agent $\alpha_j$ appears in sequence $\mathcal{S}(\psi_i)$.

We can now define a class of mechanisms that operate in the above settings. A \emph{diffusion-based task execution mechanism} specifies the following:
	$I \subseteq N$ is a set of \emph{initial nodes} to target (e.g. via advertising);
	$\rho_i$ is the \emph{payment} made to agent $\alpha_i$;
such that the following constraint is satisfied: $c |I| + \sum_{\alpha_i \in N} \rho_i \leq B$.

In words, the mechanism makes two decisions. First, it decides which nodes to target initially via advertising. Second, it decides on the payment (if any) to be made each agent. The mechanism must do this within its budget $B$.

In the DARPA Network Challenge, each $\psi_i$ represents finding a balloon, and $v(\psi_i) = 4,000$ for all $\psi_i \in \Psi$. Moreover, since the ten tasks are all identical (namely finding a balloon), $\forall \alpha_i \in N, \forall \psi_k, \psi_l \in \Psi$ we have $P(\alpha_i, \psi_k) = P(\alpha_i, \psi_l)$. That is, the success probability of a particular agent is the same for all balloons.

\begin{figure}[htbp]
  \subfigure[Example social network.]{
		\includegraphics[scale=1]{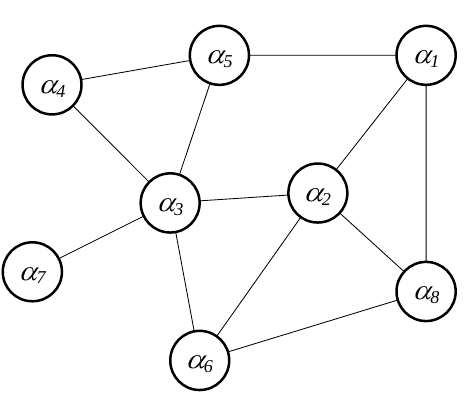} 
		\label{fig:network}
}
  \subfigure[Recruitment tree with two paths (shown in thick lines) initiated by $\alpha_1$ led to finding balloons.]{
		\includegraphics[scale=1]{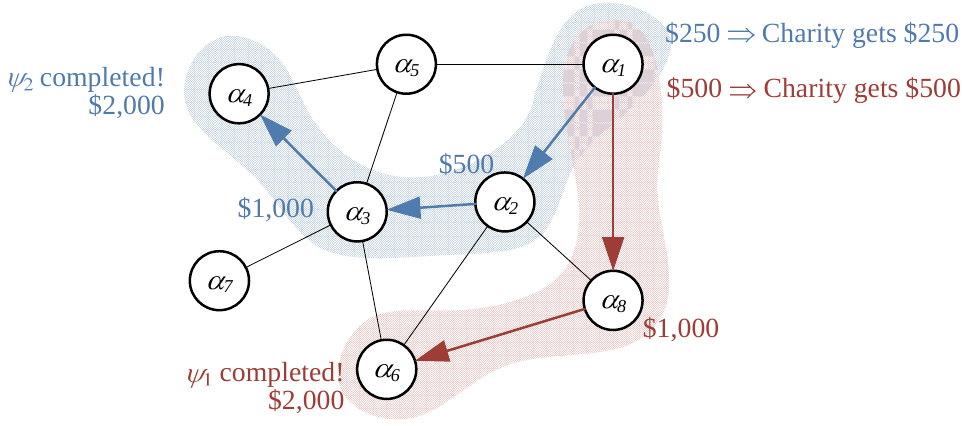} 
		\label{fig:path}
}
\caption{{\small Recursive incentive mechanism: \subref{fig:network} Suppose that in this network, agent $\alpha_1$ recruits all of his neighbors, namely $\alpha_2$, $\alpha_5$ and $\alpha_8$. Suppose that $\alpha_8$ recruits $\alpha_6$, who finds balloon $\psi_1$.
\subref{fig:path} We have a winning sequence $\mathcal{S}(\psi_1) = \langle \alpha_1, \alpha_8, \alpha_6 \rangle$ with $|\mathcal{S}(\psi_1)| = 3$. The finder receives $\rho_8 = \frac{4,000}{2^{(3-3+1)}} = 2,000$. Since $\alpha_8$ recruited $\alpha_6$, then $\rho_8 = \frac{4,000}{2^{(3-2+1)}} = 1,000$. From this sequence, $\alpha_1$ receives $\frac{4,000}{2^{(3-1+1)}} = 500$.
Likewise, looking at the left recruitment path, we have a winning sequence $\mathcal{S}(\psi_2) = \langle \alpha_1, \alpha_2, \alpha_3, \alpha_4 \rangle$ with $|\mathcal{S}(\psi_2)| = 4$. The finder receives $\rho_4 = \frac{4,000}{2^{(4-4+1)}} = 2,000$. As above, we have $\rho_3 = \frac{4,000}{2^{(4-3+1)}} = 1,000$ and $\rho_2 = \frac{4,000}{2^{(4-2+1)}} = 500$. From this sequence, $\alpha_1$ receives $\frac{4,000}{2^{(4-1+1)}} = 250$. 
Adding up its payments from the two sequences it initiated, $\alpha_1$ receives a total payment of $\rho_1 = 750$.
Assuming there are only two tasks, the surplus in this case is $S = (4,000 - 3,500) + (4,000 - 3,750) = 750$.}}
      \label{fig:example}
\end{figure}

We are now ready to define the MIT Team mechanism, referred to as a \emph{recursive incentive mechanism}.
%
Given $I$ initial targets, and assuming $v(\psi_i) = B/|\Psi|$, divide the budget $B$ such that each task $\psi_i \in \Psi$ has budget $B_i = B / |\Psi|$. If agent $j \in N$ appears in position $k$ in sequence $\mathcal{S}(\psi_i)$, then $j$ receives the following payment:
\begin{equation}
\frac{v(\psi_i)}{2^{(|\mathcal{S}(\psi_i)| - k + 1)}}
\end{equation}
Hence, the total payment received by agent $j$ is the sum of payments for all sequences in which $j$ appears:
\begin{equation}
\rho_j = \sum_{\psi_i | j \in \mathcal{S}(\psi_i)} \frac{v(\psi_i)}{2^{(|\mathcal{S}(\psi_i)| - k + 1)}}
\end{equation}
The surplus is therefore: $S = B - \sum_{\alpha_j \in N} \rho_j$.
Figure \ref{fig:example} illustrates how this mechanism works.

\section{Analysis}


The recursive incentive mechanism has a number of desirable properties. First, it is straight forward to show that the recursive incentive mechanism is never in deficit (i.e. never exceeds its budget).\footnote{See supplementary material for formal proof of this and other properties.} 

The mechanism is also resistant to certain kinds of manipulation. In particular, after being recruited by a friend, an individual has no incentive to create his own root node by visiting the Balloon Challenge Web page directly (without using the link provided by the recruiter). This follows from the fact that payment to the person finding the balloon does not depend on the length of the chain of recruiters leading to him.

On the other hand, the mechanism is not resistant to \emph{false name} attacks, which were originally identified in the context of Web-based auctions \cite{yokoo:etal:2004}. In this attack, which has been shown to plague powerful economic mechanisms such as Vickrey-Clarke-Groves \cite{yokoo:etal:2004}, an individual creates multiple false identities in order to gain an unfair advantage. In our recursive incentive mechanism, if an individual finds a balloon, and is able to create false identities, he has an incentive to recruit such identities over a chain, then declare the balloon under the last identity. This way, using $m$ false identities and recruiting them over a chain, the manipulator can obtain rewards of $\sum_{l=0}^{m} \frac{v(\psi_i)}{2^{l+1}}$. If $m$ can be arbitrarily large, the manipulator can extract the entire reward in the limit. Having said that, our data does not reveal any successful incidents of false-name attacks, which may be due to the fact that the mechanism did not operate for long enough for people to identify this potential. In practice, other measures could be put in place to prevent, minimize or detect this kind of attack, such as using certified addresses, user rating of reputation, or even criminal prosecution \cite{ebay:shilling}.

We now ask why the mechanism succeeded in mobilizing such a large number of people in a relatively short period of time. The mechanism's success can be attributed to its ability to provide incentives for individuals to \emph{both} report on found balloon locations, while simultaneously participating in the dissemination of information about the cause. Assuming that people are self-interested, when agent $\alpha_i$ becomes aware of a task $\psi \in \Psi$, it needs to select a (possibly empty) set of neighbors $T(\alpha_i) \subseteq \{\alpha_j \in N ~:~ (\alpha_i, \alpha_j) \in E \}$ to recruit (i.e. to inform them about $\psi$). The diffusion of information about the task relies crucially on such recruitment choices among agents.

One can perform this incentive analysis under two different assumptions. Under one assumption, the probability of each person finding a balloon is an independent (and very small) constant, $\forall i, k, P(\alpha_i, \psi_k) = \epsilon$, such that $n.\epsilon \leq 1$, i.e. the sum of these probabilities over the \emph{entire} population (including those not recruited) is bounded by $1$. In this case, it is trivial to show that recruiting all of one's peers is the best strategy. Without recruiting, one achieves an expected reward of $\epsilon \frac{v(\psi_i)}{2}$. With recruiting, on the other hand, one's expected rewards is $\epsilon \frac{v(\psi_i)}{2} + \sum_{j} \epsilon x_j \frac{v(\psi_i)}{2^j}$, where $x_j$ is the number of individuals at depth $j$ of the recruiter's tree. Clearly, this expected reward increases monotonically in the number of directly recruited nodes.

We can also analyze incentives under the assumption that the probability of an individual finding a balloon is uniformly distributed across the 
\emph{recruited} individuals, that is, given $R$ recruited individuals, $\forall i \in R, k \in \Psi, P(\alpha_i, \psi_k) = \frac{1}{|R|}$. Intuitively, it means that a fixed-size group of recruited individuals is guaranteed to find the balloon eventually, even if no other individuals are recruited. This assumption is realistic if the set of recruited individuals is sufficiently large (e.g. thousands). We show that, under fairly broad assumptions on the structure of the society, it is also in the best interest of each individual to recruit all their friends. In particular, we show that if no individual controls $n/2$ of the population (i.e. is able to prevent them from learning about the task), then the strategy profile in which all individuals recruit all their friends is a Nash equilibrium.\footnote{See appendix for proofs.}

The two assumptions above differ in their treatment of how the addition of a new member to the network affects the probability that each other member succeeds in finding a balloon.  The first assumption is that new members have no effect on existing members, perhaps because they are searching mutually exclusive areas, and if the new member were to find a balloon, that implies that no-one would have found that balloon in his absence.  The second assumption is that each member's probability decreases from $\frac{1}{n}$ to $\frac{1}{n+1}$, perhaps because all members share the same search space.  Unless ``network effects'' are present (e.g. working with a new member makes us both more likely to succeed than working along), these two assumptions represent best- and worst-case assumptions.  There are certainly intermediate assumptions that could be made, and the strategies that we show to be optimal at both extremes will be optimal for these intermediate assumptions as well.

\section{Empirical Data}

We have just shown that the mechanism can lead to diffusion cascades under fairly broad assumptions. The crucial question is whether this theoretical property translates to empirical success. We explore how our mechanism's performance compares with previous studies on search and recruitment in social networks.

\begin{figure}
  \subfigure[Large successful recruitment tree]{
		\includegraphics[scale=1]{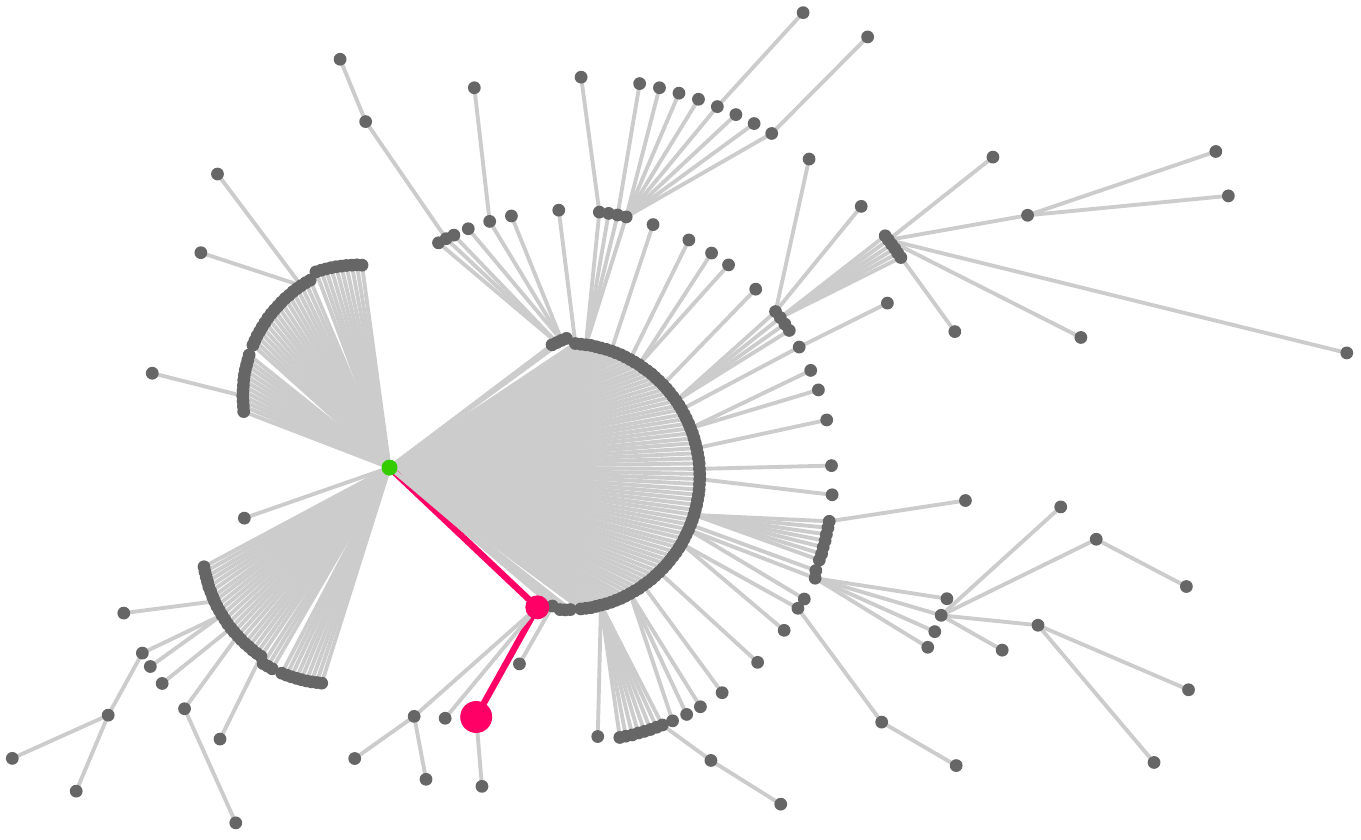} 
		\label{fig:280}
	}		\\
  \subfigure[]{
		\includegraphics[width=0.5\textwidth]{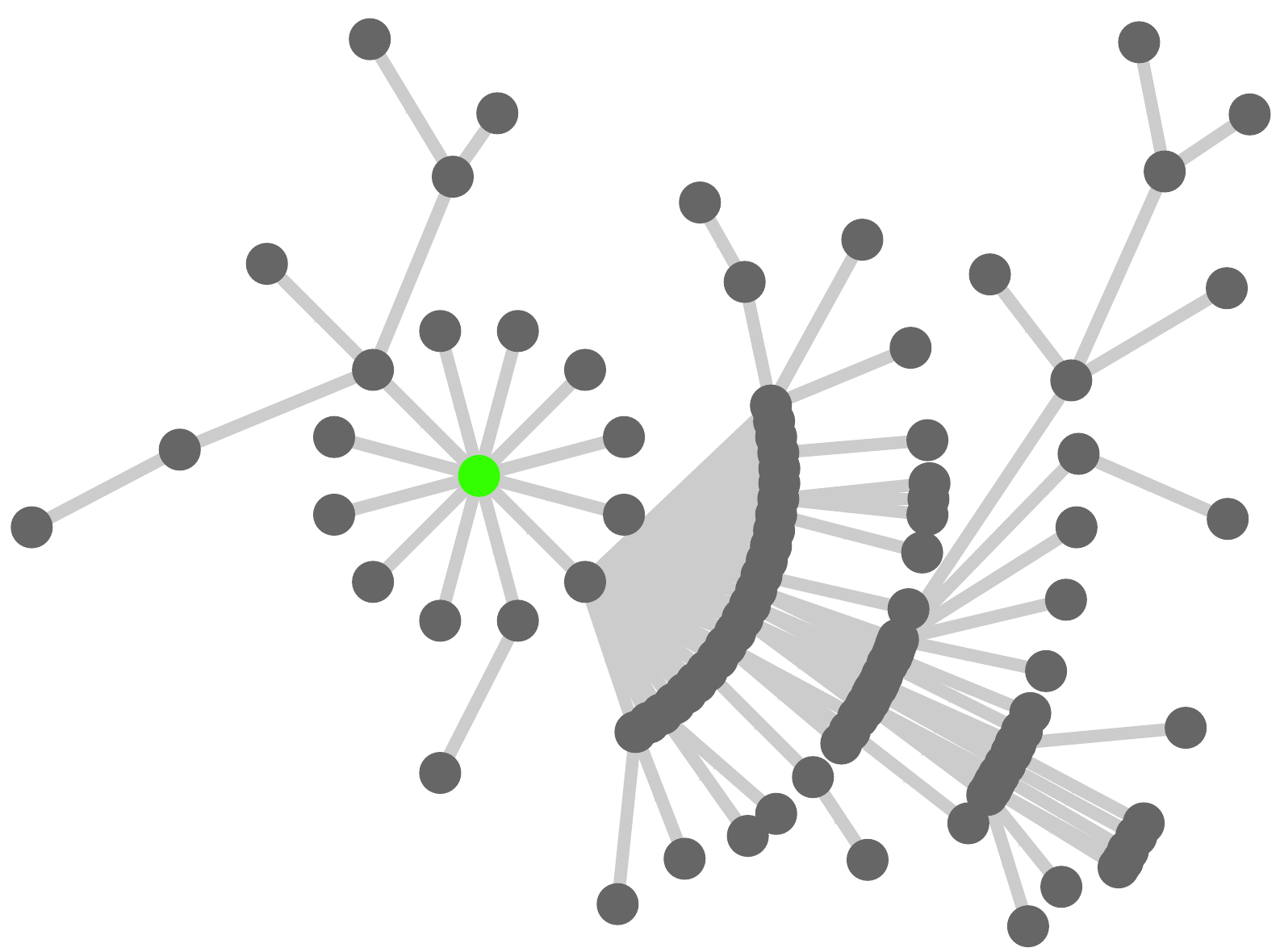} 
		\label{fig:106}
	}
  \subfigure[]{
		\includegraphics[width=0.5\textwidth]{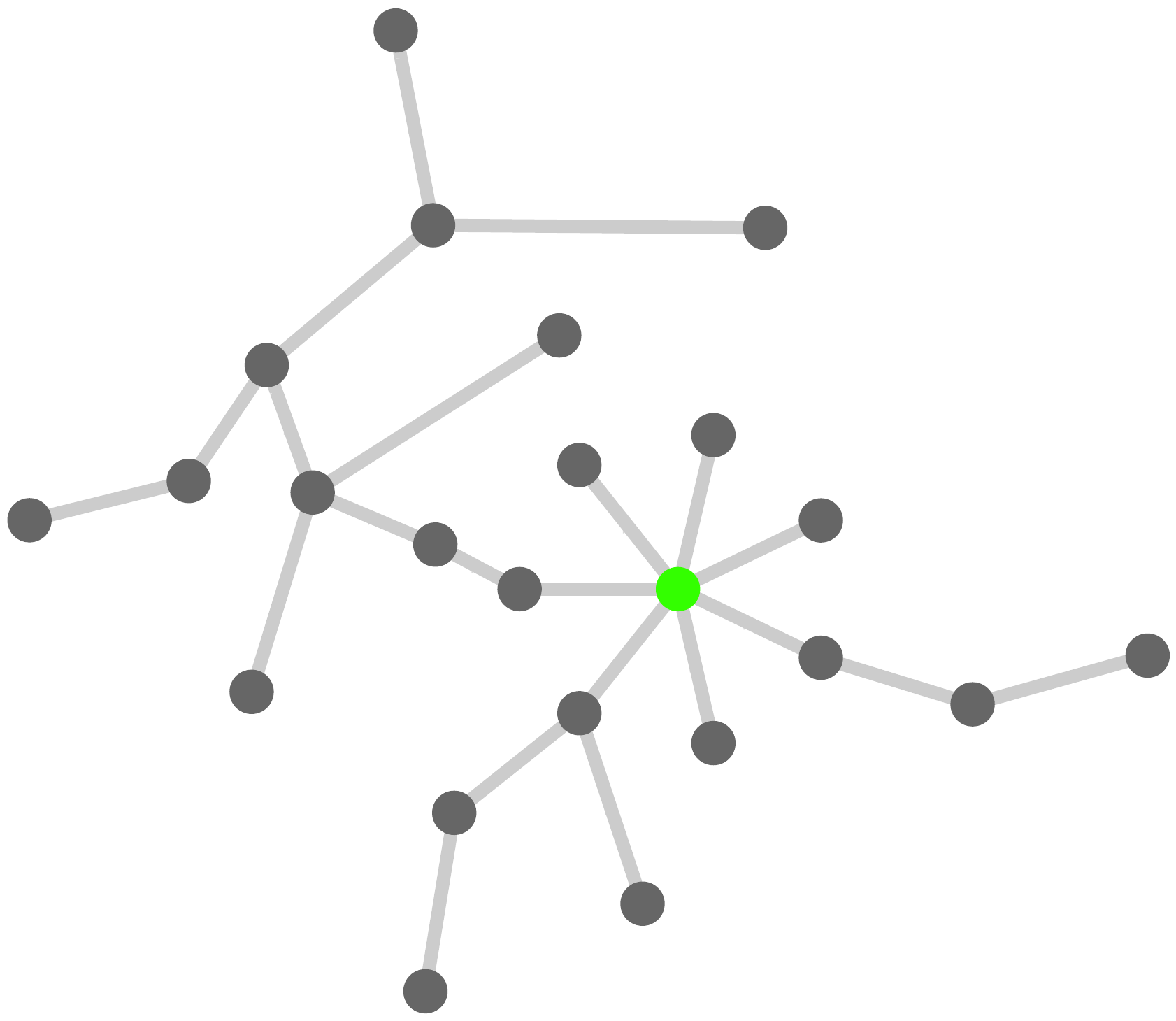}
		\label{fig:22}
	}		
\caption{{\small \subref{fig:280}  A tree with the root is shown in green, and the successful path highlighted in red. \subref{fig:106} and \subref{fig:22} Two additional networks that did not lead to balloons.}}
      \label{fig:plots:networks}
\end{figure}

\begin{figure}[htbp]
  \subfigure[Tree depth]{
		\includegraphics[scale=0.34]{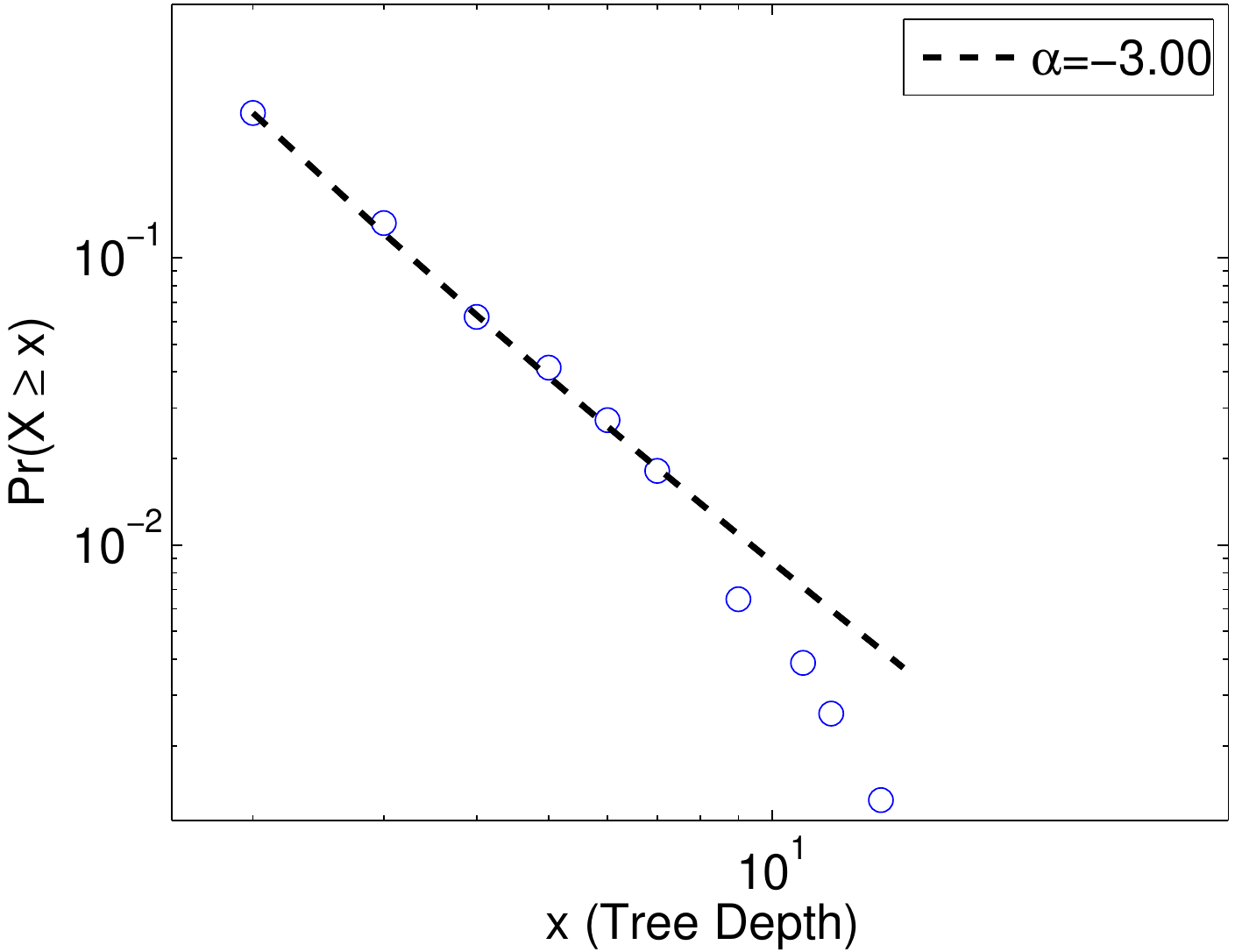}
		\label{fig:treedepth}
}
  \subfigure[Tree size]{
		\includegraphics[scale=0.34]{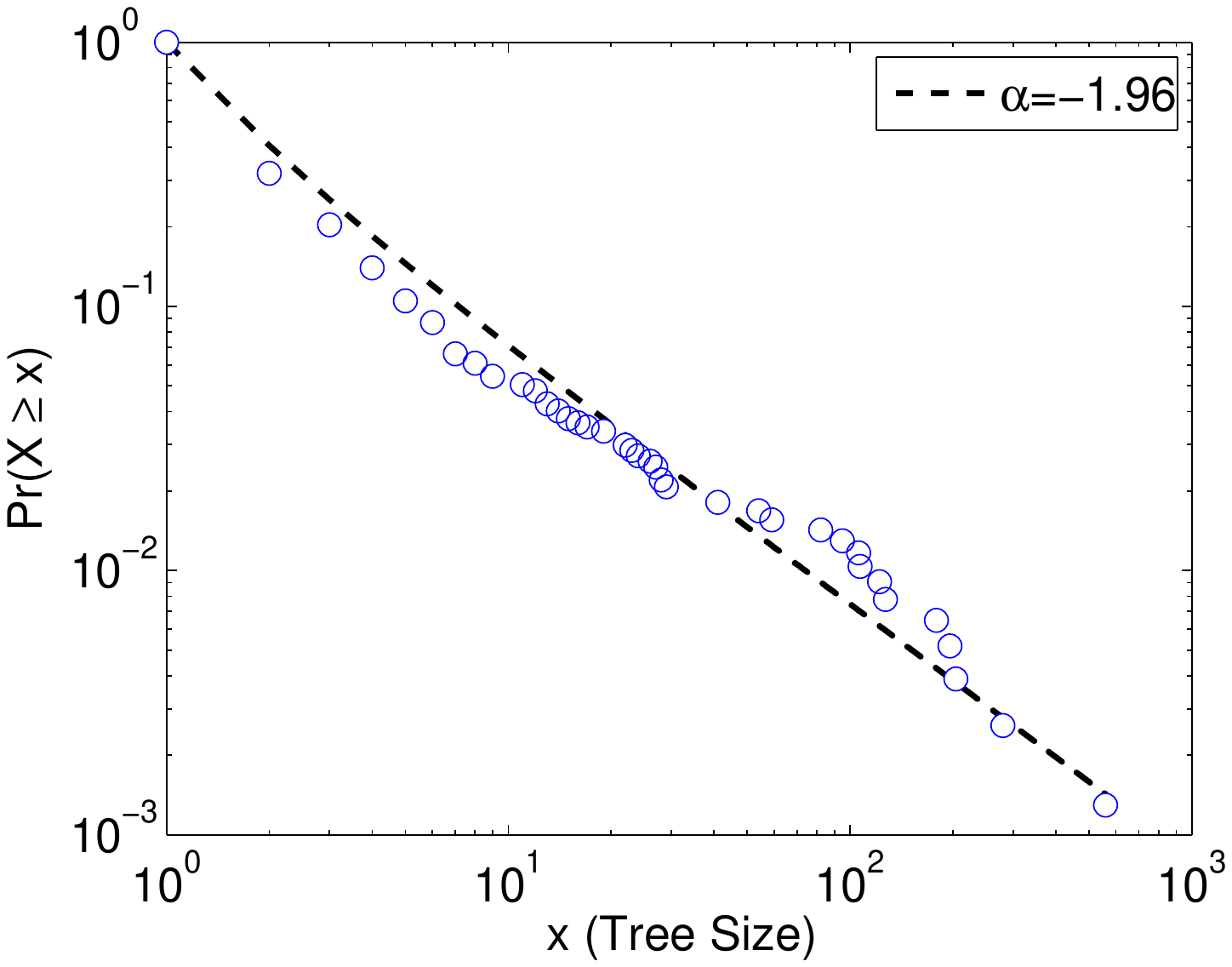} 
		\label{fig:treesize}
}
  \subfigure[Branching factor]{
		\includegraphics[scale=0.34]{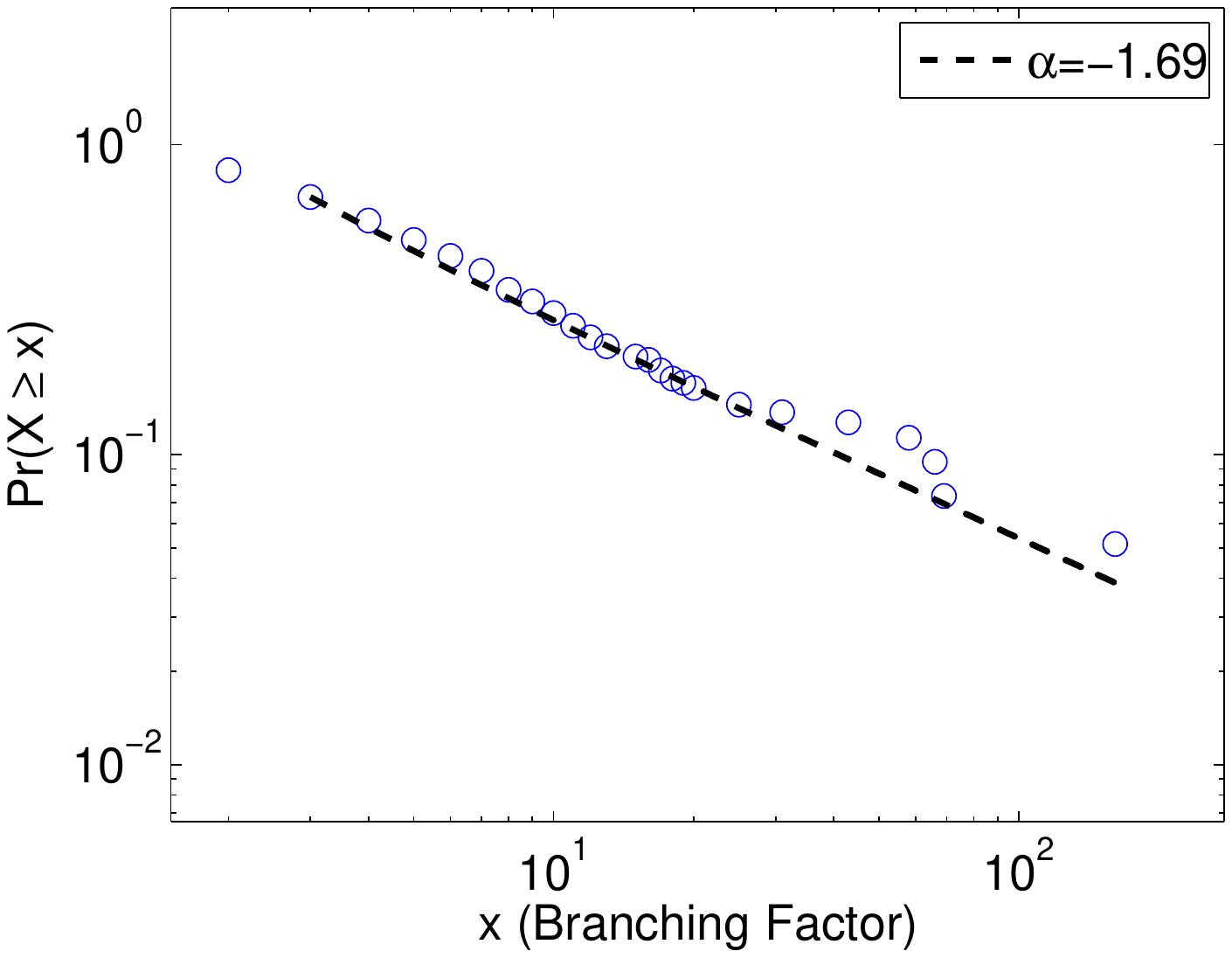} 
		\label{fig:branching}
}
\caption{{\small \subref{fig:treedepth} Distribution of tree depth on a log-log scale with a power law fit. \subref{fig:treesize} Distribution of tree size on a log-log scale with a power law fit. \subref{fig:branching} Distribution of the branching factor on a log-log scale with a power law fit.}}
      \label{fig:plots}
\end{figure}

One measure of success is the \emph{size of the cascades}, both in terms of \emph{number of nodes}, as well as \emph{depth}. Results vary in existing literature. In a study of the spread of online newsletter subscriptions, in which individuals were rewarded for recommending the newsletter to their friends, the $7,188$ cascades varied in size between $2$ and $146$ nodes, with a maximum depth of $8$ steps \cite{iribarren:moro:2009}, over a time span of three months.\footnote{Esteban Moro, personal communication.} In our data, if we ignore the MIT root node, there are $845$ trees recruited within only three days. The largest tree contained $602$ nodes, and the deepest tree was $14$ levels deep. Figure \ref{fig:plots:networks} shows three actual trees, with Figure \ref{fig:280} highlighting a successful path. Figure \ref{fig:treedepth} shows the distribution of tree/cascade depth, which follows a power law. Furthermore, Figure \ref{fig:treesize} shows a power-law distribution of tree/cascade size with exponent $-1.96$, as predicted by models of information avalanches on sparse networks \cite{watts:2002}.




Previous empirical studies reported significant \emph{attrition rates} (aka \emph{discard rate}), which measures the percentage of nodes that terminate the diffusion process. For example, in a study of email-based global search for $18$ target persons, attrition rate varied between $60-68\%$ in $17$ out of the $18$ searches performed \cite{dodds:etal:2003}. It has been argued that the ``\textit{lack of interest or incentive, not difficulty, was the main reason for chain termination}'' \cite{dodds:etal:2003}. In another study of the diffusion of online recommendations, an attrition rate of $91.21\%$ was repoted \emph{despite} providing incentives to participants by offering them a chance in a lottery \cite{iribarren:moro:2009}. In the DARPA Network Challenge, if we ignore isolated single nodes, our mechanism achieves a significantly lower attrition rate of $56\%$. 

Another measure of performance for social mobilization processes is the \emph{branching factor} (also known as the \emph{reproductive number}), which is the number of people recruited by each individual. Previous empirical studies reported diverse, though mostly low, observations. In a study of the spread of support for online petitions, dissemination was very narrow, with more than $90\%$ of nodes having exactly one child \cite{liben-nowell:kleinberg:2008}, which others have attributed to a selection bias, observing only large diffusions \cite{golub:jackson:2010}. In our data, the average branching factor was $0.93$ if we exclude single-node trees ($0.80$ if we include single-node trees). As shown in \ref{fig:branching} shows, the branching factor follows a power-law distribution, suggesting that certain individuals played an important role in dissemination by recruiting a very large number of people (E.g. see Figure \ref{fig:280}). Our data also compares very favorably with the newsletter subscription experiment mentioned above, in which spreaders invited an average of $2.96$ individuals, but were only able to cause $0.26$ individuals to sign up on average \cite{iribarren:moro:2009}.

\begin{figure}[htbp]
  \subfigure[Recruitment over time]{
		\includegraphics[scale=0.34]{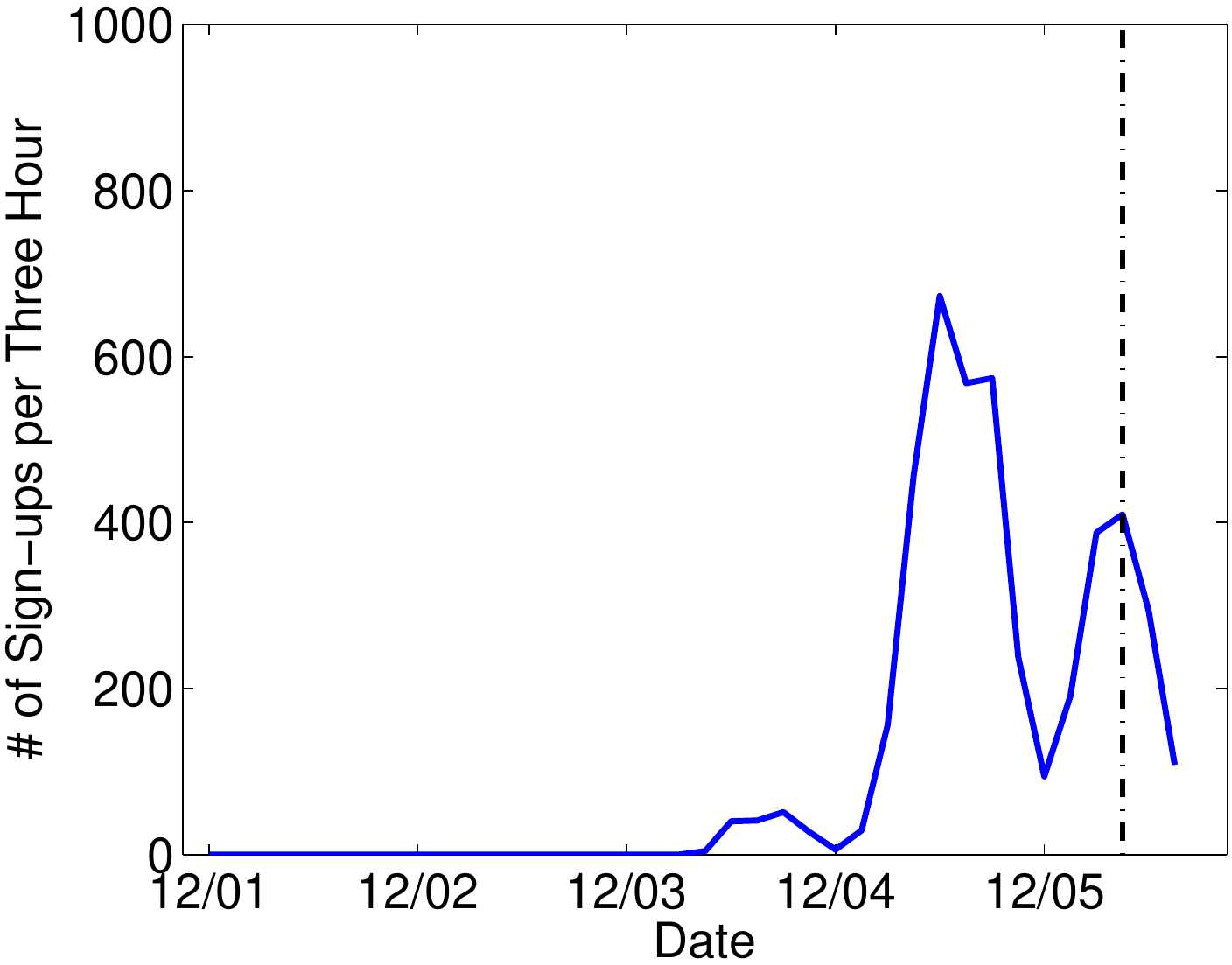} 
		\label{fig:numsignupNonaccum}
	}		
  \subfigure[Team size over time]{
		\includegraphics[scale=0.34]{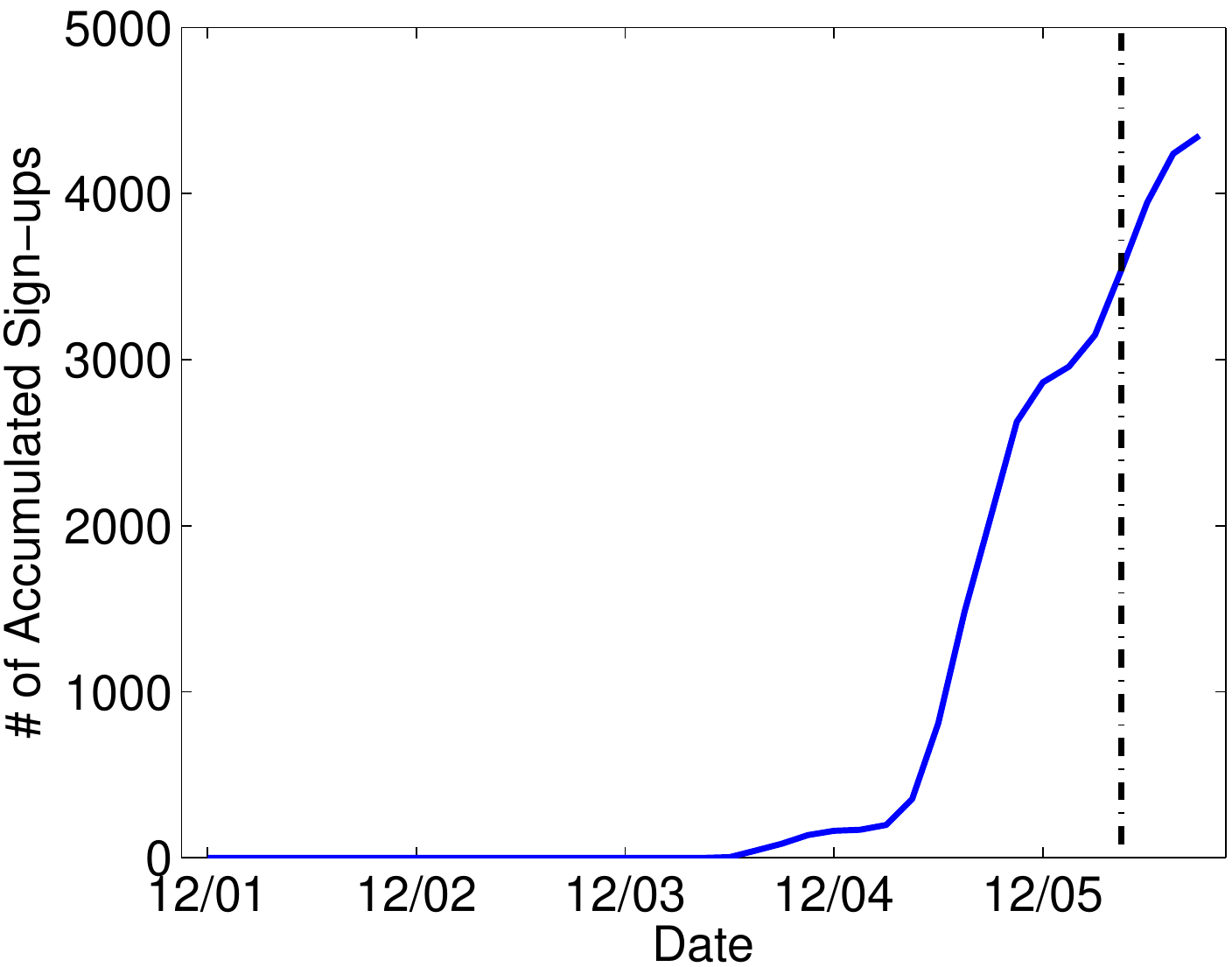} 
		\label{fig:numsignup}
	}		
  \subfigure[Inter-signup time]{
		\includegraphics[scale=0.34]{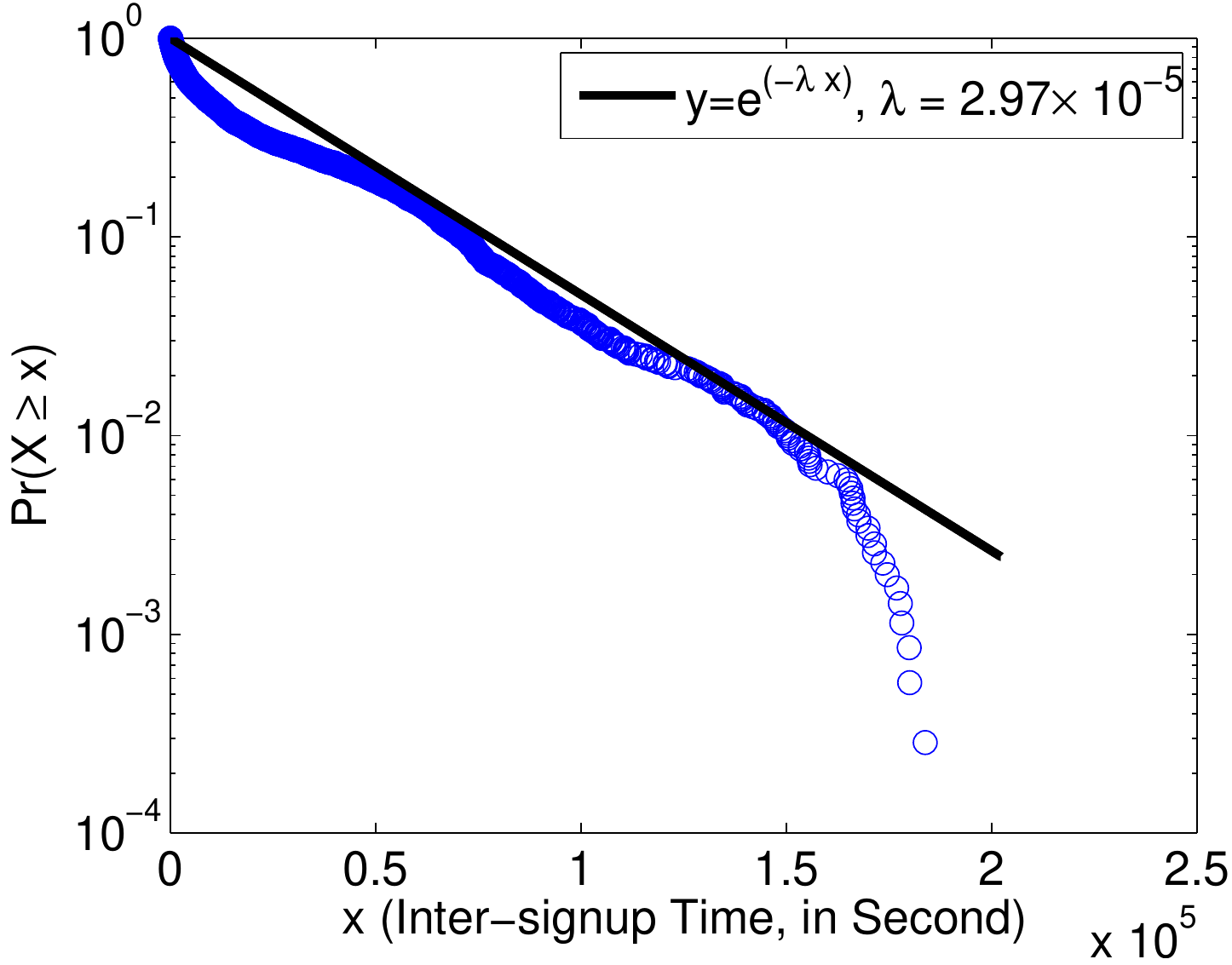} 
		\label{fig:signupDelay}
	}
\caption{{\small \subref{fig:numsignupNonaccum}  Number of people recruited over time up to the winner announcement. The dotted line marks the time the balloons were launched into their positions by DARPA. \subref{fig:numsignup} Cumulative number of people recruited over time. \subref{fig:signupDelay} Complementary cumulative distribution of the inter-signup time on a semi-log scale with an exponential fit. Observe the larger-than-exponential drop off at the end of the graph, due to the time-critical nature of the task.}}
      \label{fig:plots:time}
\end{figure}

An interesting aspect of our data is the dynamics of the diffusion process. Figures \ref{fig:numsignupNonaccum} and \ref{fig:numsignup} show the dynamics of recruitment over time, highlighting two bursts of day-time recruitment activities on Friday and Saturday just before DARPA launched the balloons into their locations. In contrast with the newsletter subscription experiment \cite{iribarren:moro:2009}, in which diffusion experienced a continuous decay, these bursts enabled our mechanism to amass a large number of people quickly.

Moreover, in the newsletter subscription experiment \cite{iribarren:moro:2009}, the dynamics of diffusion were slow, which was attributed to a heterogeneous, non-Poissonian distribution of individuals' response time. Interestingly, we observe an exponential distribution of inter-signup time (See Figure \ref{fig:signupDelay}).\footnote{Our data does not include the time stamp of sending out invitations, but we are able to measure the intervals between actual signup events between a parent and its children in the trees.} This contrasts with the empirically observed power-law distribution of inter-response time in human activity \cite{barabasi:2005,malmgren:etal:2009} and information cascades \cite{iribarren:moro:2009}. Ongoing initiatives that utilize our approach could determine whether this deviation is due to the incentive mechanism.\footnote{Similar mechanisms, inspired by our approach, are being used to spread petitions for fighting world hunger (\url{www.1billionhungry.org}), in games of cooperation and prediction \url{http://brsts.com/}, and for marketing campaigns (\url{https://10.thinkworld.com.cn/}).}


\section{Conclusion}

From an \emph{observational perspective}, previous studies have shown that the success of information cascades on social networks is affected by various factors, such as the percentage of targeted individuals \cite{watts:peretti:2007}, the heterogeneity in response time \cite{barabasi:2010,iribarren:moro:2009,liben-nowell:kleinberg:2008}, the types of social ties used in the spread \cite{dodds:etal:2003}, the heterogeneity in response thresholds among nodes \cite{watts:2002}, and the density of the network \cite{watts:2002,liben-nowell:kleinberg:2008}. However, from the \emph{perspective of an incentive designer} seeking large-scale social mobilization, the problem boils down to two questions: (i) which individuals to target directly? (ii) what incentives to provide in order to encourage participation? While others have addressed the first question \cite{domingos:richardson:2001,kempe:etal:2003,leskovec:etal:2007,kitsak:etal:2010}, here we addressed the question of incentives, which has not received much attention in the literature, as pointed out by Dodds et al \cite{dodds:etal:2003},  until the DARPA Network Challenge. In particular, our mechanism \emph{simultaneously} provides incentives for participation and for recruiting more individuals to the cause. This mechanism is already being used in different contexts, such as social mobilization to fight world hunger, in games of cooperation and prediction, and for marketing campaigns. 

We believe that it is not a coincidence that the winning strategy in the DARPA Network Challenge was one that exploited ideas from both incentive design \cite{maskin:2008} and computational social science \cite{lazer:etal:2009}. After all, people are self-interested individuals, but also embedded within social networks. It is hoped that this paper will stimulate theoretical and empirical efforts to devise incentive mechanisms for a variety of challenging, time-critical social mobilization problems.


\bibliographystyle{Science}
\bibliography{mybib}

\newpage

\appendix 

\section{Supporting Online Material: Formal Proofs}

\subsection{Mechanism Always within Budget}

\begin{proposition}
The recursive incentive mechanism is never in deficit (i.e. never exceeds its budget).

\begin{proof}
Recall that each sub-task $\psi_i$ is allocated an equal share of $B_i = B/ |\Psi|$ budget. Hence, it suffices to show that the payment for any arbitrary task $\psi_i$ is bounded by $B_i$. Let $\mathcal{S}(\psi_i) = \langle a_1, \dots, a_r \rangle$ be the (finite) sequence leading to the successful completion of $\psi_i$. 
We need to show that the total payment made to all agents in sequence $\mathcal{S}(\psi_i)$ within budget, that is, we need to show that:
$$\sum_{k=1}^{r} \frac{B_i}{2^{(r - k + 1)}} \leq B_i ~~~~\textit{or equivalently we need to show that}~~~~ \sum_{k=1}^{r} \frac{1}{2^{(r - k + 1)}} \leq 1$$
We can easily see that: $\sum_{k=1}^{r} \frac{1}{2^{(r - k + 1)}} = \sum_{k=1}^{r} (\frac{1}{2})^{(r - k + 1)} = \sum_{k=1}^{r} \frac{1}{2} \times (\frac{1}{2})^{(r - k)}$

Defining $i = r-k$, we can rewrite: 
\begin{eqnarray*}
\sum_{k=1}^{r} 0.5 (\text{\textonehalf})^{(r - k)} & = & 
0.5 (\text{\textonehalf})^{(r - 1)} + 0.5 (\text{\textonehalf})^{(r - 2)} + \dots 0.5 (\text{\textonehalf})^{(r - r)} \\
& = & 0.5 (\text{\textonehalf})^{(r - 1)} + 0.5 (\text{\textonehalf})^{(r - 2)} + \dots 0.5 (\text{\textonehalf})^{0} \\
& = & \sum_{i=0}^{r-1} 0.5 (\text{\textonehalf})^{i}
\end{eqnarray*}
This is a finite geometric series, with a well-known closed form:
$$
\sum_{i=0}^{r-1} 0.5 (\text{\textonehalf})^{i} = 0.5 \frac{1 - (\text{\textonehalf})^{(r-1)+1}}{1 - \text{\textonehalf}} 
																							 = 1 - (\text{\textonehalf})^r 
																							 = \frac{2^{r} - 1}{2^{r}} \leq 1 ~~~~~(\text{for}~~ r \geq 1)$$
\end{proof}
\end{proposition}

\subsection{Incentives With Uniform Success Probability Among Recruited Individuals}

\subsubsection{All-or-None Recruitment on Fixed-Forest Social Networks}

We consider the case in which the social network takes the form of a forest of rooted trees, and the roots of these trees form the set of initially-recruited nodes $I$.

Given this forest $F$, which contains a total of $n$ nodes, each node chooses whether or not to recruit all of its children.  This induces a ``recruited subforest'' $F'$ of size $n'$, consisting of all nodes which can trace
a path of recruitment to a root node of $F$.

\begin{figure}[h]
\begin{center}
\ovalbox{
\begin{tikzpicture}[level/.style={sibling distance=25mm/#1},edge from parent/.style={draw,thick},every node/.style={draw,black}]
  \path node at (0,0) [shape=circle, draw=blue!50, fill=blue!20, thick, minimum size=10mm] {$R_1$}
		child {node [shape=circle, draw=blue!50, fill=blue!20, thick, minimum size=10mm] {$a_1$} edge from parent[red]
			child {node [shape=circle, draw=blue!50, fill=blue!20, thick, minimum size=10mm] {$c_1$} edge from parent[black]}
			child {node [shape=circle, draw=blue!50, fill=blue!20, thick, minimum size=10mm] {$d_1$} edge from parent[black]}
		}
		child {node [shape=circle, draw=blue!50, fill=blue!20, thick, minimum size=10mm] {$b_1$} edge from parent[red]
			child {node [shape=circle, draw=blue!50, fill=blue!20, thick, minimum size=10mm] {$e_1$}}
			child {node [shape=circle, draw=blue!50, fill=blue!20, thick, minimum size=10mm] {$f_1$}}
		}
	node at (5,0) [shape=circle, draw=blue!50, fill=blue!20, thick, minimum size=10mm] {$R_2$}
		child {node [shape=circle, draw=blue!50, fill=blue!20, thick, minimum size=10mm] {$a_2$}
			child {node [shape=circle, draw=blue!50, fill=blue!20, thick, minimum size=10mm] {$c_2$} edge from parent[red]}
			child {node [shape=circle, draw=blue!50, fill=blue!20, thick, minimum size=10mm] {$d_2$} edge from parent[red]}
		}
		child {node [shape=circle, draw=blue!50, fill=blue!20, thick, minimum size=10mm] {$b_2$}
			child {node [shape=circle, draw=blue!50, fill=blue!20, thick, minimum size=10mm] {$e_2$}}
			child {node [shape=circle, draw=blue!50, fill=blue!20, thick, minimum size=10mm] {$f_2$}}
		}
	node at (10,0) [shape=circle, draw=red!50, fill=red!20, thick, minimum size=10mm] {$R_1$} 
		child {node [shape=circle, draw=red!50, fill=red!20, thick, minimum size=10mm] {$a_1$} edge from parent[red]}
		child {node [shape=circle, draw=red!50, fill=red!20, thick, minimum size=10mm] {$b_1$} edge from parent[red]
			child {node [shape=circle, draw=red!50, fill=red!20, thick, minimum size=10mm] {$e_1$}}
			child {node [shape=circle, draw=red!50, fill=red!20, thick, minimum size=10mm] {$f_1$}}
		}
	node at (13,0) [shape=circle, draw=red!50, fill=red!20, thick, minimum size=10mm] {$R_2$};
		
\end{tikzpicture}
}
\end{center}
\caption {Nodes $R_1$, $b_1$, and $a_2$ choose to recruit; the rest not.  The recruited subforest $F'$ is shown in red.  Note that $a_2$'s choice to recruit is rendered moot by $R_2$'s choice not to recruit.}
\end{figure}
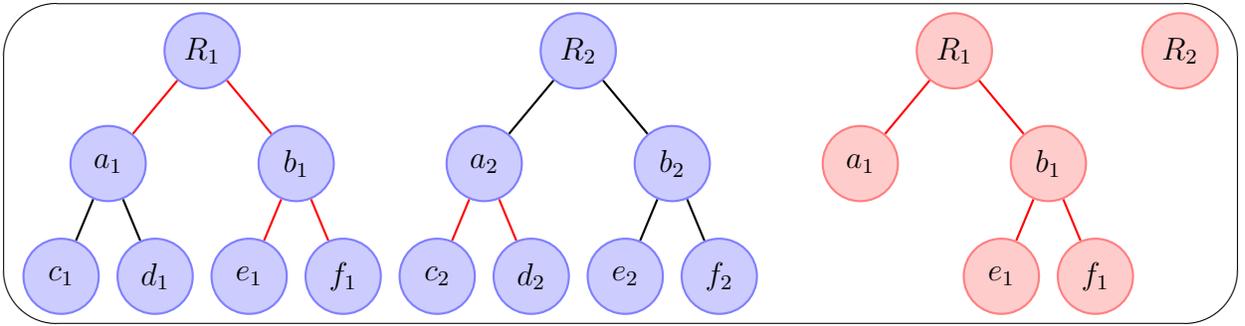

For each node in the recruited subforest, this results in an expected payment based solely on the shape of its descendent recruited subtree.  For each node, we can characterize this shape with an ordered tuple $X = \langle x_1, x_2, x_3, \ldots 0\rangle$, representing the number of children, grandchildren, great-grandchildren, etc. (i.e. in the example, $R_1$'s tuple would be $\langle 2, 2, 0\rangle$, and the tuple of any leaf node is $\langle 0 \rangle$).  Given such a tuple, the expected payment to a node is $$U(X) = \frac{1+\sum_i \frac{x_i}{2^i}}{n'},$$ where $n'$ is the number of nodes in the recruited subforest.

Given the set of choices (recruit all children or recruit no children) made by each node, this function $U(X)$ is a payout function which defines a normal-form game played by all non-leaf nodes in the original forest.

\subsubsection{Game Definition}

We demonstrate the definition of the game by example, recreating the ``prisoner's dilemma'' using a 5-node forest.

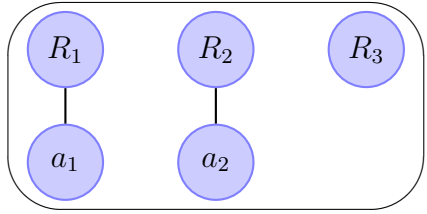
\begin{figure}[t]
\begin{center}
\ovalbox{
\begin{tikzpicture}[level/.style={sibling distance=25mm/#1},edge from parent/.style={draw,thick},every node/.style={draw,black}]
  \path node at (0,0) [shape=circle, draw=blue!50, fill=blue!20, thick, minimum size=10mm] {$R_1$}
		child {node [shape=circle, draw=blue!50, fill=blue!20, thick, minimum size=10mm] {$a_1$}}
	node at (2,0) [shape=circle, draw=blue!50, fill=blue!20, thick, minimum size=10mm] {$R_2$}
		child {node [shape=circle, draw=blue!50, fill=blue!20, thick, minimum size=10mm] {$a_2$}}
	node at (4,0) [shape=circle, draw=blue!50, fill=blue!20, thick, minimum size=10mm] {$R_3$};
		
\end{tikzpicture}
}
\end{center}
\caption {The game played by $R_1$ and $R_2$ is equivalent to the ``prisoner's dilemma.''}
\label{fig:pd}
\end{figure}

Consider the forest $F$ shown in Figure~\ref{fig:pd}.  There are two players, $R_1$ and $R_2$, each of which has the option to recruit
a single child or not.  If neither recruits, both receive an expected payment of $\frac{1}{3}$.  If one recruits but the other does not, the
recruiter has an expected payment of $\frac{1+\frac{1}{2}}{4} = \frac{3}{8}$, while the other has an expected payment of $\frac{1}{4}$.  If both recruit, both have an expected payment of $\frac{1+\frac{1}{2}}{5} = \frac{3}{10}$.  This gives a payment matrix approximated by:

\begin{game}{2}{2} & $N$ & $Y$\\ $N$ &$.33,.33$ &$.25,.37$\\ $Y$ &$.37,.25$ &$.3,.3$ \end{game} \\

Clearly, choosing to recruit is a strictly dominant strategy for each player, so the only Nash equilibrium that both players recruit -- even though this is Pareto inefficient.

\subsection{Nash Equilibria of Larger Forests}

\begin{lemma}All nodes choosing to recruit is a Nash equilibrium for the game defined by a forest $F$ if and only if each node prefers recruitment over non-recruitment, predicated on all other nodes choosing to recruit.

\begin{proof}

Consider a game in which all actors have two options: ``recruit all'' or ``recruit none.''  For any given agent $a$, let all other agents choose ``recruit all,'' and consider $a$'s optimal strategy.  If choosing ``recruit all'' is optimal for $a$, then no agent can benefit by deviating from a strategy of ``recruit all,'' if all other agents choose ``recruit all.''  This, by definition, makes the uniform choice to ``recruit all'' a Nash equilibrium.

\end{proof}
\end{lemma}

\begin{theorem}A node $a$ will prefer recruitment to non-recruitment predicated on all other nodes choosing to recruit if and only if sufficiently many
nodes in the forest $F$ are not descendants of $a$.

\begin{proof}
For a node $a$ in a forest $F$ of size $n$, let the tuple $X = \{x_1, x_2, x_3, \ldots\}$ be defined as the number of children, grand-children, great-grand-children, etc. of node $a$.  If $F$ is finite, each $x_i$ is finite and there exists some $j$ such that $x_i = 0$ for all $i > j$.  Let $k$ be the number of nodes in $F$ that are not descendants of $a$, noting that $k = n - \sum_i{x_i}$.  Since we assume all nodes other than $a$ choose to recruit, the expected payment received by $a$ if $a$ chooses to recruit is $\frac{1}{k}$.  If $a$ does choose to recruit, then $a$ will receive expected payment $\frac{1+\sum_i \frac{x_i}{2^i}}{n} = \frac{1 + \sum_i \frac{x_i}{2^i}}{k + \sum_i{x_i}}$.  $a$ will find it preferable to recruit if and only if $\frac{1 + \sum_i \frac{x_i}{2^i}}{k + \sum_i{x_i}} > \frac{1}{k}$, or, equivalently, when $k > \frac{\sum_i{x_i}}{\sum_i\frac{x_i}{2^i}}$.

\end{proof}
\end{theorem}

\begin{corollary}In any forest $F$ of size $n$ for which no tree contains more than $\frac{n}{2}$ nodes, all nodes choosing to recruit is a Nash equilibrium.

\begin{proof}

Consider forest $F$ with $n$ nodes, and a node $a$ which has $m$ descendants, taking a shape described by a tuple $X = \{x_1, x_2, x_3, \ldots\}$.  We have that $a$ will choose to recruit predicated on all other nodes recruiting if and only if $n-m > \frac{m}{\sum_i\frac{x_i}{2^i}}$.  We note that the definition of $X$ yields that no non-zero value can follow a zero value (i.e. one must have grand-children in order to have great-grand-children).  It follows that, if we fix $m$, the setting of $X$ which maximizes $\frac{m}{\sum_i\frac{x_i}{2^i}}$ is $X = \{\overbrace{1, 1, \ldots, 1}^m, 0, 0, \ldots\}$, which gives $\sum_i\frac{x_i}{2^i} < 1$ for any value of $m$.  Thus, $a$ will choose to recruit if (but not only if) $n-m > m$.  This condition holds for all nodes if and only if no tree in $F$ contains more than $\frac{n}{2}$ nodes.  In this case, all nodes will choose to recruit predicated on all other nodes recruiting, so all nodes choosing to recruit is a Nash equilibrium.

\end{proof}
\end{corollary}

\subsection{Selective Recruitment on Fixed-Forest Networks}

We now consider the same social graph structure, but allow a node to selectively recruit any subset of its children.

\begin{definition}[Weight]
We define the weight of a node $a$, $W_a$, as the sum of the rewards that would be received by $a$ in the event that each of its descendants were to complete the task.  We note the following properties of $W_a$

\begin{itemize}

\item If $a$ is a leaf, then $W_a = 1$.

\item If $a$ has children $c_1, c_2, \ldots$ with weights $W_{c_1}, W_{c_2}, \ldots$, then $W_a = 1 + \frac{1}{2}\sum_i W_{c_i}$.

\item If node $a$ has descendants described by shape $X = <x_1, x_2, \ldots, 0>$, then $W_a = 1 + \sum_i \frac{x_i}{2^i}$.

\item In a forest with $n$ nodes, the expected payment to node $a$ is $U(a) = \frac{W_a}{n}$.

\end{itemize}
\end{definition}

\begin{lemma} A node $a$ will prefer recruitment of all children to non-recruitment of any child predicated on all other nodes choosing to recruit if and only if the weights of $a$'s children are sufficiently large relative to the number of their descendants.

\begin{proof}

Consider a node $a$ with children $c_1, c_2, \ldots, c_m$, and let all other nodes choose to recruit all of their children.  Let $|c_1|, |c_2|, \ldots, |c_m|$ be the number of descendants of each child of $a$, and let $k$ be the number of nodes in the forest that are not descendants of $a$.  $a$ can choose to recruit each child independently.  When $a$ recruits no children, its expected payment is $\frac{1}{k}$.  When $a$ recruits all of its children, its expected payment is $\frac{1 + \frac{1}{2}\sum_i W_{c_i}}{k + \sum_i |c_i|}$.  For a given child $c_x$, the expected gain by recruiting $c_x$ is monotonically non-increasing over the set of other children recruited (i.e. if it is advantageous to recruit $c_x$ when also recruiting all other children, it will be advantageous to recruit $c_x$ when recruiting any subset of the other children).  For a given child $c_x$, it is advantageous to recruit $c_x$ if and only if $\frac{1 + \frac{1}{2}\sum_{i \neq x} W_{c_i}}{k + \sum_{i \neq x} |c_i|} < \frac{1 + \frac{1}{2}\sum_i W_{c_i}}{k + \sum_i |c_i|}$.  $\frac{1 + \frac{1}{2}\sum_{i \neq x} W_{c_i}}{k + \sum_{i \neq x} |c_i|}$ is maximized in the case where all children $c_{i \neq x}$ have no children of their own, and in that case equal to $\frac{1 + \frac{1}{2}(m-1)}{k+{m-1}}$.  Thus, we can guarantee that the inequality holds so long as $\frac{W_{c_x}}{|c_x|} > \frac{1 + \frac{1}{2}(m-1)}{k+{m-1}}$.

\end{proof}
\end{lemma}

\begin{theorem} A node $a$ will prefer recruitment of all children to non-recruitment of any child predicated on all other nodes choosing to recruit if sufficiently many nodes in the forest $F$ are not descendants of $a$.

\begin{proof}

Consider a node $c_x$, and the inequality $\frac{W_{c_x}}{|c_x|} > \frac{1 + \frac{1}{2}(m-1)}{k+{m-1}}$.  As before, the left side of the inequality is minimized when the children of $c_i$ form a chain with no branching.  This chain has weight $\sum_{i=0}^{|c_x|} \frac{1}{2^i}< 2$, which bounds the left side of the inequality by $\frac{2}{|c_x|}$.  In a forest in which no tree contains more than $\frac{n}{4}$ nodes, $|c_x|$ is bounded above by $\frac{n}{4}$, and $k$ is bounded below by $\frac{3n}{4}$.  $m$ can only take values in the range $0 \leq m \leq \frac{n}{4}$, and for any such value, the inequality holds.  Thus, so long as at least $\frac{3n}{4}$ nodes in $F$ are not descendants of $a$, $a$ will choose to recruit all of its children.

\end{proof}
\end{theorem}

\subsection{Recruitment on Graphs}

We consider now the case in which the social graph is not a forest, but is instead a general graph.  In this case, the mechanism of recruitment itself plays a non-trivial role, since it is possible for a node to be recruited by two different potential parents, and must choose between them.  There is significant literature on diffusion processes on graphs, and wide varieties of such processes are seen in practice.  We will not investigate the properties of specific diffusion mechanisms, but instead we will define a property of a diffusion mechanism that guarantees that recruitment is Nash.

\begin{definition}[Monotonic Diffusion]
Consider a diffusion process on a social graph, and a set of seed nodes $R_1, R_2, \ldots, R_n$.  Let $|R_1|, |R_2|, \ldots, |R_n|$ be the number of nodes whose recruitment leads back to $R_1, R_2, \ldots, R_n$, respectively.  We call the diffusion process monotonic if removing a seed node $R_x$ causes the sizes of $|R_1|, |R_2|, \ldots, |R_n|$ to either increase or stay constant (i.e. if $R_x$ does not participate, this does not cause another seed node to recruit fewer children).
\end{definition}

Monotonicity holds for most ``well-behaved'' diffusion processes, but is notably violated by various ``complex contagion'' processes in which, for example, a node adopts after receiving two signals.

\begin{theorem} If no node can expect to recruit more than half of the social network and diffusion is monotonic, then all nodes recruiting is a Nash equilibrium.

\begin{proof}

Consider a node $a$, which can choose whether or not to recruit, and suppose all other nodes recruit.  Consider the case in which $a$ recruits, and this results in no tree in the induced forest containing more than half of the recruited nodes.  Suppose it were the case that if $a$ were to not choose recruitment, then all nodes that would have been recruited by $a$ would end up un-recruited, instead.  In this case, the graph reduces to the same fixed forest we analyzed previously.  Suppose instead that some of these nodes end up recruited by a different node.  In this case, not recruiting is strictly less desirable, since the size of the network grows without any increase in potential payout.  Hence, it follows from the previous analysis that recruiting is more desirable in either case.  If diffusion is monotonic, the two cases considered are collectively exhaustive, so recruiting is always the more desirable option.

\end{proof}
\end{theorem} 


\end{document}